\documentclass[twocolumn, apex]{jjap3}
\usepackage{newtxtext,newtxmath}
\usepackage{braket}
\usepackage{bm}

\title{Determination of the electron trap level in Fe-doped GaN by phonon-assisted conduction phenomenon}
\author{Hiroki Fukuda$^{1}$, Akira Nagakubo$^{1}$, Shigeyoshi Usami$^{1}$, Masashi Ikeda$^{1}$, Masayuki Imanishi$^{1}$, Masashi Yoshimura$^{1}$, Yusuke Mori$^{1}$, Kanta Adachi$^{2,3}$, and Hirotsugu Ogi$^{1}$\thanks{E-mail: ogi@prec.eng.osaka-u.ac.jp}}
\inst{$^1$Graduate School of Engineering, Osaka University, Suita, Osaka 565-0871, Japan\\
$^2$Department of Systems Innovation Engineering, Iwate University, 4-3-5 Ueda, Morioka, Iwate, 020-8551, Japan\\
$^3$Department of Earth Sciences, University of Cambridge, Downing Street, Cambridge CB2 3EQ, UK
}
\abst{We acoustically measured the energy level for thermally activated conduction (TAC) in high-resistivity Fe-doped GaN using the non-contacting antenna-transmission acoustic-resonance method.  The acoustic attenuation takes a maximum at a specific temperature, where the TAC is accelerated with the help of phonon energy. The Debye type relaxation is thus observed for acoustic attenuation, and its activation energy (0.54$\pm$0.04 eV) was determined with attenuation measurements at various frequencies and temperatures. This value agrees with the E3 level in GaN, indicating that thermally associated conduction originates from the E3 trap level.  We also measured the five independent elastic constants at high temperatures.}

\begin{document}
\maketitle

Excellent electronic properties of GaN, such as the high break-down electric field and high mobility, make it an attractive material for next-generation high-power and high-frequency devices like AlGaN/GaN high electron mobility transistors (HEMTs)\cite{HEMT-Shinohara, GaN-Review}. However, their performance is deteriorated by the charge trapping\cite{CC, CC-Surface, CC-Yang, CC-Deep-Fe}. High-resistivity GaN doped with Fe or C is used in the buffer layers of HEMTs to suppress the leakage current, and some studies have reported that the high-resistivity layer traps the electrons\cite{CC-Deep-Trap-1, Trap-0.57eV-Fe}. These trapped electrons can be thermally activated to cause the hopping conduction\cite{HC-DCLook}. Therefore, it is important to understand the mechanisms related to the thermally activated conduction (TAC) in Fe-doped GaN. Another important issue is to measure the elastic constants at high temperatures because HEMTs are expected to be used at elevated temperatures\cite{HEMT-HighTemp-Daumiller, HEMT-HighTemp-Egawa}. A lattice mismatch between AlGaN and GaN produces the interfacial strain, which creates the two-dimensional electron gas through the piezoelectrocity\cite{HEMT-strain-Yalamarthy, HEMT-HighTemp-strain}. Therefore, to understand the carrier transport properties at high temperatures, it is necessary to measure the elastic constants of GaN at high temperatures. However, there are no experimental reports about the elastic constants of GaN beyond 100 $^{\circ}$C. 

We previously proposed a methodology to observe TAC through ultrasonic attenuation using the non-contacting antenna-transmission acoustic-resonance (ATAR) method\cite{HC-Ogi}. The trapped carriers jump to the other sites with the help of phonon energy, and this behavior can be expressed by the Debye type relaxation.  It is thus possible to determine the relaxation time by measuring the temperature dependence of the ultrasonic attenuation. The deep level transient spectroscopy (DLTS), widely known as a method of detecting deep levels in semiconductors, is difficult to apply to high resistivity materials\cite{HR-DLTS}: The DLTS signal becomes nearly the noise level at Fe concentrations of 10$^{17}$ cm$^{-3}$ or higher, commonly used in HEMTs. In contrast, the ATAR method is  applicable to high-resistivity piezoelectric materials such as GaN with Fe concentrations of $\sim$10$^{19}$ cm$^{-3}$. 

The ATAR method, however, relies on the piezoelectricity of the material, which deteriorates with the progress of TAC, and we cannot use such a vibrational mode that is highly affected by TAC because of disappearance of the resonant mode. In this study, we find that TAC is suppressed in quasi-shear resonance modes, and we overcome this problem using them. Using this technique, we measure internal friction of various quasi-shear resonance modes and determine the activation energy from the Arrhenius plot to be 0.54 eV. We also measure the elastic constants of $C_{44}$ and $C_{33}$ at high temperatures.

We used a rectangular-parallelepiped Fe-doped wurtzite GaN grown by the hydride-vapor-phase-epitaxy method. The resistivity at room temperature and Fe concentration are 2 $\times$ 10$^8$ $\Omega$ cm and 8 $\times$ 10$^{18}$ cm$^{-3}$, respectively. Fe acts as a deep accepter so that the inherent piezoelectricity of GaN appears, allowing us to vibrate the specimen using an alternating electric field. The in-plane specimen dimension is 3.503$\times$2.995 mm$^2$, and the thickness (along the $c$ axis) is 0.413 mm. 

\begin{figure}[tb]
  \begin{center}
  \includegraphics[width=8cm]{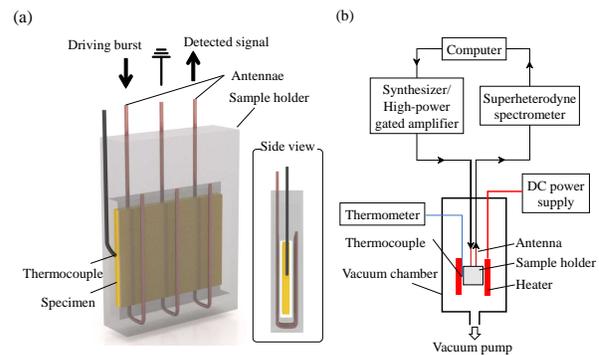}
  \caption{Schematic diagrams of (a) the sample holder and (b) the measurement system.}
  \label{Fig-system}
  \end{center}
\end{figure}

We have developed the ATAR system\cite{HC-Ogi, ATAR-Nakamura} to study TAC of Fe-doped GaN. Figure \ref{Fig-system} shows the measurement setup. We developed the sample holder with alumina cement as shown in Fig. \ref{Fig-system}(a), where three Ni-Cr alloy wires (usable at high temperatures) are embedded for generation, detection, and ground antennae, so as to surround the specimen inside it. The sample space of the holder is a little larger than the specimen, and no external force is applied to the specimen except for gravity. Note that this is highly important for achieving accurate internal-friction measurement. We vibrate the specimen through the inverse-piezoelectric effect by applying tone-burst voltage to the generation antenna. After the excitation, the vibration of the specimen generates the electric field through the piezoelectric effect, which is detected by the detection antenna. The acoustic measurement is thus performed in a non-contacting manner.  We changed the specimen temperature by inserting the sample holder into a heater. We set a thermocouple beside the specimen to measure the specimen temperature. We placed them into a vacuum chamber as shown in Fig. \ref{Fig-system}(b).

We applied the tone bursts to the generation antenna through a high-power gated amplifier to vibrate the specimen and detected the reverberation signal by the detection antenna, which entered a superheterodyne spectrometer to extract the phase and amplitude of the driving-frequency component. The experiment takes two steps.  First, we performed the frequency-scan experiment to obtain a resonant spectrum, from which we determined the resonant frequency.  Second, we measured the attenuation curve by driving the generation antenna with the measured resonant frequency and determined the attenuation coefficient \cite{HC-Ogi, ATAR-Nakamura} and then internal friction $Q^{-1}$.  This attenuation measurement was used for resonant modes below 15 MHz.  For higher frequency modes, we determined internal friction from the bandwidth of the resonant peak.


We observed TAC caused by the piezoelectric polarization change induced by the mechanical vibration of the specimen. The rate of TAC ($\tau^{-1}$) exponentially increases with increasing temperature, and when it matches the acoustic frequency, a part of the acoustic energy is transferred to the carrier movement. Thus, internal friction shows a peak when the specimen temperature is changed. We assumed the Debye-type relaxation for this phenomenon\cite{Hutson-loss, HC-Ogi},
\begin{align}
  Q^{-1} = \frac{e^2}{C\varepsilon} \frac{\omega \tau}{1+(\omega \tau)^2}. \label{eq-Qinv}
\end{align}
Here, $\omega$ is the angular frequency, and $e$, $C$, and $\varepsilon$ are the effective piezoelectric, elastic, and dielectric constants dominating the resonance mode, respectively. $\tau$ is the relaxation time for TAC and takes the form of $\tau \propto \exp(E/k_{\rm B} T)$, where $E$ denotes the activation energy for TAC. Therefore, we can determine the relaxation time from the internal-friction peak at $\omega \tau$=1. 

\begin{figure}[tb]
  \begin{center}
  \includegraphics[width=8cm]{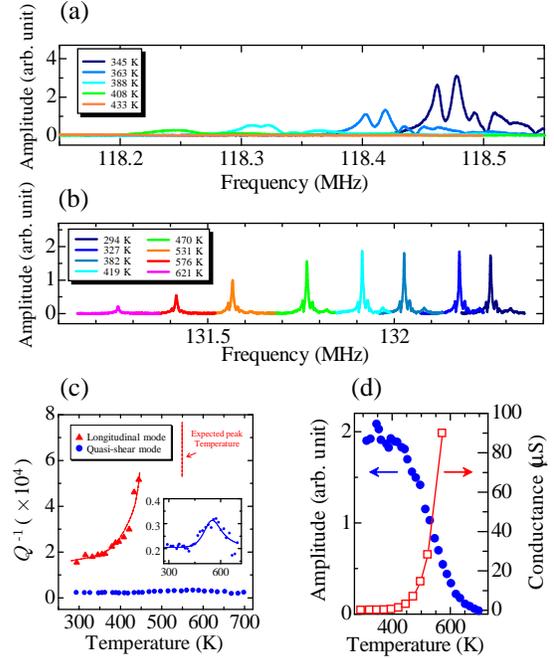}
  \caption{Measured resonance spectra of the (a) longitudinal mode and (b) quasi-shear resonance mode at various temperatures.  (c) Measured internal friction for the longitudinal mode (red triangles) and quasi-shear mode (blue circles). The red dashed curve is a guide for eyes, and the red dashed vertical line indicates the expected internal-friction peak of the longitudinal mode. The blue solid line is fitted Eq. (\ref{eq-Qinv}). (d) Temperature dependence of conductance (red open squares) and resonance amplitude of the quasi-shear resonance mode (blue closed circles).}
  \label{Fig-LongShear}
  \end{center}
\end{figure}

Hutson {\it et al.} showed that internal friction of elastic waves in piezoelectric semiconductors is affected by piezoelectrically stiffened phenomenon\cite{Hutson-loss}.There are three independent piezoelectric coefficients for GaN ($e_{15}$, $e_{31}$, and $e_{33}$), and the absolute value of $e_{33}$ is significantly larger than those of the others \cite{Adachi-piezo}. One may, therefore, consider that using a resonant mode that is largely affected by $e_{33}$ would be appropriate for the study of TAC from Eq. (1).  The through-thickness longitudinal-wave resonant mode would be the best one because its apparent elastic constant $C'_{33}$ is the most remarkably stiffened by $e_{33}$ as
\begin{align}
  C'_{33} &= \frac{4d^2\rho f^2}{n^2} = C_{33} + \frac{e_{33}^2}{\varepsilon_{33}}, \label{eq-C33}
\end{align}
where $d, \rho, f$, and $n$ denote the thickness, mass density, resonant frequency, and the resonance mode number; and $C_{33}$ denotes the longitudinal-wave elastic constant along the $c$ axis.  However, the ATAR method requires the piezoelectricity for the non-contact vibration excitation and detection, and it fails to observe the resonance mode when TAC occurs to a large extent.  For example, Figure \ref{Fig-LongShear}(a) shows that resonant spectra at various temperatures for a longitudinal-wave resonant mode, where the resonant peak cannot be detected at temperatures higher than $\sim$430 K.  The internal-friction peak is accordingly unmeasurable as shown in Fig. \ref{Fig-LongShear}(c), in which the internal-friction peak is expected to appear at 550 K. In contrast, the shear-plane wave resonance is unaffected by TAC because the elastic constant $C'_{44}$ for this resonance is free from the piezoelectric coefficients:  
\begin{align}
C'_{44} &= \frac{4d^2\rho f^2}{n^2} = C_{44}, \label{eq-C44}
\end{align}
where $C_{44}$ is the shear elastic constant. Such a resonant mode can be excited through $e_{15}$ with the ATAR method and be measurable at higher temperatures because of suppressed TAC.  However, it is insensitive to TAC.

Namely, (i) we need to use the ATAR method for accurate internal-friction measurement.   (ii) We need to use a $C_{33}$ related resonant mode for the study of TAC, but (iii) we have to suppress TAC for higher temperature measurement.  We here propose to use the quasi-shear resonant modes to overcome this labyrinthine problem.  Since the actual specimen has a finite size, the pure shear resonances do not occur. We calculated resonance frequencies and contributions of individual elastic constants in detail (See Supplementary Material) to identify the quasi-shear resonant modes.  For example, the resonant mode near 132 MHz (corresponding to the 27th shear-plane wave resonance) consists of $\sim$95\%$C_{44}$ and $\sim$2\%$C_{33}$ contributions, and others.  This mode can be observable over 600 K as shown in Fig. \ref{Fig-LongShear}(b), and its internal friction is measurable up to 700 K (Fig. \ref{Fig-LongShear}(c)).  Owing to the high accuracy of the attenuation measurement in the ATAR method, the internal-friction peak is clearly observed even with the small contribution of $C_{33}$, although the resonance-peak amplitude of the quasi-shear mode decreases with increasing temperature because thermally excited free carriers deteriorate the piezoelectricity. (The increase in the specimen conductance corresponds to the decrease in the resonance amplitude as shown in Fig. \ref{Fig-LongShear}(d)).  

\begin{figure}[tb]
  \begin{center}
  \includegraphics[width=8cm]{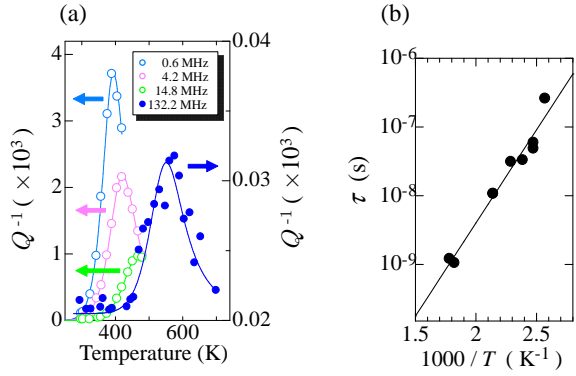}
  \caption{(a) Temperature dependence of internal friction of various quasi-shear resonance modes. Solid lines represent the theoretical internal friction based on the Debye-type relaxation form of Eq. (\ref{eq-Qinv}). (b) Relationship between the relaxation time and temperature. The activation energy of 0.54 eV is obtained from its slope.}
  \label{Fig-QinvArrhenius}
  \end{center}
\end{figure}

We thus measured the temperature dependence of internal friction for the quasi-shear resonance modes in a wide frequency range between 0.6 and 150 MHz as shown in Fig. \ref{Fig-QinvArrhenius}(a) and determined the relaxation time $\tau$ at each temperature as shown in Fig. \ref{Fig-QinvArrhenius}(b).  We then obtained the activation energy for TAC from the Arrhenius plot to be $0.54 \pm 0.04$ eV. This value is significantly different from that determined in the previous study \cite{HC-Ogi}(0.23 eV) with a much narrower frequency range (0.5--15 MHz). We consider the value determined in this study is more reliable because of the ten-times higher frequency range used. 

The acoustical method used here for determining the activation energy has advantages over other methods like the temperature-dependent conduction measurement\cite{Imp-conduction-Mott,Imp-conduction-Fritzsche,Energy-Hache}. It can separately detect a single relaxation phenomenon and determine the corresponding activation energy because each internal-friction peak reflects one relaxation mechanism. On the other hand, in the conduction measurement, for example, the overall temperature behavior of conduction is measured even when multi relaxation mechanisms participate in the conduction, and it is difficult to separately determine each activation energy.  

There have been many experimental and theoretical reports of trap levels in Fe-doped GaN. Kordo\v{s} {\it et al.} \cite{Energy-Cond-Hall-APL-2004} provided 0.58--0.6 eV for a deep Fe acceptor using the conductivity and Hall-effect measurements. Polyakov {\it et al.} \cite{Energy-PICTS-APL-2003} showed $\sim$0.5 eV for the depth for trapping the dominant electrons pinning the Fermi level with the photoinduced current transient spectroscopy. Density-functional-theory calculations indicated that an Fe atom substitutes the Ga site and forms an energy level at $E_C - 0.5$ eV\cite{Energy-First-1, Energy-First-2}, where $E_C$ denotes the condition band minimum. Recently, Horita {\it et al.} have reported that there is a strong correlation between the Fe concentration and the trapped-electron concentration, and they suggest that the origin of the trap level is the Fe substitution\cite{Energy-Horita}. From these previous reports, the activation energy obtained in this study can be identified as that for a TAC phenomenon caused by hopping of electrons in the deep trap level formed by the substitution of the Fe atom to the conduction band minimum. 

\begin{figure}[t]
  \begin{center}
  \includegraphics[width=8cm]{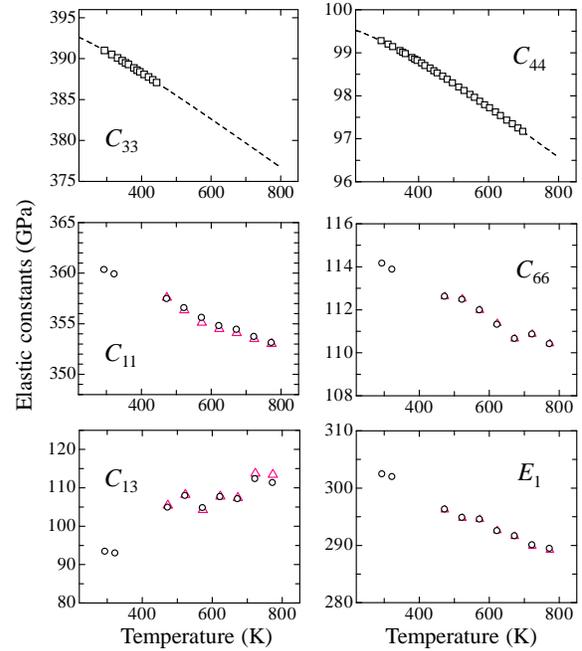}
  \caption{Temperature dependence of the elastic constants of GaN. The black squares in $C_{33}$ and $C_{44}$ are obtained by ATAR, and dashed lines indicate fitted theory based on Varshni's equation\cite{Varshni}. The black circles show the in-plane elastic constants ($C_{11}$, $C_{66}$, $C_{13}$, and $E_{1}$) inversely determined using $C_{33}$ and $C_{44}$ values from Varshni's theory\cite{Varshni}. The red triangles show those determined by fixing the $C_{33}$ and $C_{44}$ values at their maximum temperatures.}
  \label{Fig-EC}
  \end{center}
\end{figure}

\renewcommand{\arraystretch}{1.1}
\begin{table}[t]
  \centering
  \caption{Elastic constants $C_{ij}$ (GPa) (upper row) and the temperature coefficients $(dC_{ij}/dT)/C_{ij}$ (ppm/K) (lower row) around room temperature of wurtzite GaN and other materials.}
    \begin{tabular}{lccccccc}
    \Hline
        & $C_{33}$ & $C_{44}$ & $C_{11}$  & $C_{13}$ & $C_{66}$ & $E_{1}$ &  \\
    \hline
    GaN  & 391.0 & 99.3  & 360.3  & 93.4  & 114.1 & 302.4 &  \\
          & -72 & -54 & -43    & 418 & -71 & -93 &  \\
    Ti\cite{EC-Fisher} & 180.7 & 46.7  & 162.4   & 69.0  & 35.2  & 104.4 &  \\
          & -219  & -418  & -345     & 65    & -1033 & -875  &  \\
    Zr\cite{EC-Fisher} & 164.8 & 32.0  & 143.4   & 65.3  & 35.3  & 99.6  &  \\
          & -219 & -422  & -334     & 29    & -967  & -794  &  \\
    Hf\cite{EC-Fisher} & 196.9 & 55.7  & 181.1   & 66.1  & 66.1  & 139.9 &  \\
          & -154  & -314 & -201     & 17    & -470  & -362  &  \\
    Mg\cite{EC-Mg} & 61.60 & 16.40 & 59.40  & 21.44 & 16.90 & 45.60 &  \\
          & -341  & -599  & -302     & -69   & -477  & -426  &  \\
    Y\cite{EC-Y} & 76.9  & 24.3  & 77.9    & 21    & 24.71 & 65.0  &  \\
          & -146  & -406  & -241     & 571   & -340  & -350  &  \\
    $\alpha$-Al$_2$O$_3$\cite{EC-Al2O3} & 501.1 & 147.2 & 497.3  & 117.0 & 167.2   & 425.0 \\
          & -61   & -137  & -67       & -103  & -97      & -84 \\
    \Hline
    \end{tabular}%
  \label{Table-EC}%
\end{table}%
\renewcommand{\arraystretch}{1}

Next, we describe the measurement of the elastic constants at elevated temperatures. For this purpose, we combined the ATAR method and the tripod-type resonant ultrasound spec- troscopy (RUS).  The latter uses a tripod consisting of two needle-type piezoelectric transducers and a needle-type thermocouple, on which the specimen is set without any coupling material.\cite{RUS-Adachi}  Vibrations of the specimen are generated and detected by the two needle-type piezoelectric transducers. In principle, the RUS method allows the determination of all of the independent elastic constants of the specimen inversely by comparing the measured and calculated resonance frequencies.  However, the monocrystal GaN specimen usually has a very large aspect ratio, and it is difficult to accurately determine the out-of-plane elastic constants $C_{33}$ and $C_{44}$, because their contributions to the resonance frequencies are negligible compared with others. We therefore measured $C_{33}$ and $C_{44}$ separately by the ATAR method using Eqs. (\ref{eq-C33}) and (\ref{eq-C44}) with the 12th resonance mode ($\sim$120 MHz) for $C_{33}$ and the 27th resonance mode ($\sim$130 MHz) for $C_{44}$, where reported piezoelectric constant\cite{e33} and dielectric constant\cite{epsilon33} were involved.  We then determined the other three elastic constants ($C_{11}$,  $C_{13}$, and $C_{66}$) by performing the inverse calculation using known $C_{33}$ and $C_{44}$ values. Variations of thickness and mass density due to thermal expansion were considered using reported lattice parameters\cite{Temperature-LC}. The mass density used was 6080 kg/m$^3$ at room temperature\cite{EC-Adachi}, which is close to the theoretical value of 6084 kg/m$^3$. We computed resonance frequencies using the Ritz method with basis functions consisting of the product of normalized Legendre polynomials\cite{Ohno, Ogi}.  In this study, we involved the Legendre polynomials with orders of $N=$24 or smaller. We failed to measure the resonance frequency by the ATAR method above 444 K for $C_{33}$ and 700 K for $C_{44}$ because of TAC, and we do not have values for them in these temperature ranges. Therefore, we used extrapolated $C_{33}$ and $C_{44}$ values using Varshni's theory\cite{Varshni} (broken lines in Fig. 4). One may doubt the accuracy of the in-plane elastic constants at higher temperatures because of the lack of measurements for $C_{33}$ and $C_{44}$.  However, the contributions of the in-plane elastic constants are considerably large, and the accuracy is not so sensitive to the ambiguity of $C_{33}$ and $C_{44}$.  This is demonstrated by red marks in Fig. 4, which were deduced using the constant $C_{33}$ and $C_{44}$ values at their highest temperatures: Even with those inappropriate values, the resultant in-plane elastic constants are nearly the same as those determined with $C_{33}$ and $C_{44}$ values estimated by Varshni's theory. 

Figure \ref{Fig-EC} and Table \ref{Table-EC} show the determined elastic constants. (Note that we also failed to measure resonance frequencies by RUS around 400 K because of strong attenuation induced by TAC. ) We here compare the temperature dependence of the elastic constants of GaN to that of other hexagonal materials and $\alpha$-Al$_2$O$_3$. Their room temperature values and temperature coefficients are shown in Table \ref{Table-EC}. We expect that the temperature coefficient for $C_{33}$ is smaller than that of $C_{44}$, because the larger the elastic constant, the more insensitive it is to the temperature change due to the strong interatomic bonds. This holds for most hexagonal materials as shown in Table \ref{Table-EC}.  However, GaN fails to obey this trend. One possible explanation is the overestimation of the piezoelectric stiffening effect at elevated temperatures. As shown in Fig. \ref{Fig-LongShear}(d), piezoelectricity deteriorates with increasing temperature. Therefore, the apparent $C_{33}$ values at elevated temperatures become lower than the actual values.  Further study is needed to elucidate this unusual temperature dependence of the elastic constants of GaN.

In conclusion, we measured TAC in Fe-doped GaN using the non-contacting ATAR method. Piezoelectric disappearance caused by TAC is suppressed in the quasi-shear resonance modes, allowing us to measure the internal friction peaks at sufficiently high temperatures. We determined the activation energy of TAC to be 0.54 $\pm$ 0.04 eV, which agrees with the E3 level in GaN. We also measured the elastic constants at elevated temperatures. The absolute value of the temperature coefficient of $C_{33}$ is significantly higher than that of $C_{44}$, which is the opposite behavior for other hexagonal elements. The deterioration of the piezoelectricity at high temperatures will lead to the overestimation of the temperature coefficient.

\acknowledgment
This study was supported by JSPS KAKENHI Grant Nos. JP19H00862 and 20K21144.

\end{document}